\documentclass[prd,twocolumn,preprintnumbers,superscriptaddress,nofootinbib]{revtex4}
\pdfoutput=1

\usepackage{amsmath}
\usepackage{amssymb}
\usepackage{graphicx}
\usepackage{xspace}
\usepackage{epsfig}
\usepackage{units}
\usepackage{slashed}
\usepackage[hyperfootnotes=false]{hyperref}
\usepackage{gensymb}
\usepackage{cancel}
\usepackage{multirow}
\usepackage{float}
\usepackage[normalem]{ulem}

\newcommand{\eezh}{e^+ e^- \to ZH}
\allowdisplaybreaks

\begin{document}

\title{Two-Loop Electroweak Corrections with Fermion Loops to \boldmath $e^+ e^- \to ZH$}

\author{Ayres Freitas}
 \email{afreitas@pitt.edu}
\author{Qian Song}
 \email{qis26@pitt.edu}
\affiliation{%
Pittsburgh Particle-physics Astro-physics \& Cosmology Center(PITT-PACC)\\
Department of Physics \& Astronomy, University of Pittsburgh, Pittsburgh, PA 15260, USA
}%

\begin{abstract}
We present a complete calculation of the next-to-next-to-leading (NNLO) electroweak corrections involving closed fermion loops to the $e^+ e^- \to ZH$ process. This has been achieved by using a semi-numerical technique for the two-loop vertex and box diagrams, which is based on Feynman parameters and dispersion relations for one of the two sub-loops. UV divergences are treated with suitable subtraction terms. Numerical results for the unpolarized differential and integrated cross-section at center-of-mass energy $\sqrt{s}=240~\text{GeV}$ are provided. Combining the NNLO contributions with one and two closed fermions loops, they are found to increase the NLO cross-section by 0.7\%. 
\end{abstract}

\maketitle

\section{Introduction}

After the discovery of the Higgs boson \cite{higgs11,higgs12} at the Large Hadron Collider (LHC) in 2012, it will be crucial to perform precision studies of its properties, in order to understand the details of the mechanism of electroweak symmetry breaking and search for signs of new physics beyond the Standard Model (SM).Possible deviations of Higgs couplings from the SM expectations may appear at the per-cent level in a wide range of models \cite{Englert:2014uua}. 

For this purpose, several proposals have been made for so-called $e^+e^-$ Higgs
factories: the International Linear Collider (ILC) \cite{ilc1,ilc2}, the Future Circular Collider (FCC-ee)
\cite{fccee}, and the Circular Electron-Positron Collider (CEPC) \cite{cepc}. Those colliders are intended to operate at center-of-mass energies of 240--250~GeV, in which the Higgsstrahlung process, $\eezh$, becomes the dominant Higgs production channel. As a result of clean environment and high luminosity, the cross section for $ZH$ production is expected to be measured with a precision of about 1.2$\%$ at ILC, 0.4$\%$ at FCC-ee, and 0.5$\%$ at CEPC.

To extract the coupling between Higgs and Z boson, theoretical predictions for the process \(\eezh \) are necessary, and the precision should be at least of the same order as the experimental one. Within the SM, leading order (LO) \cite{lo} and next-to-leading order (NLO) corrections have been known since a long time for unpolarized beams \cite{nlo1,nlo2,nlo3}, and more recently for polarized beams \cite{Bondarenko:2018sgg}. The effects of multiple collinear photon emission in the initial state, which are enhanced by powers of $\log(s/m_e^2)$, can be taken into account with Monte-Carlo \cite{isr1} or structure function \cite{isr2} methods. Other higher-order corrections are more challenging to compute.
The mixed electroweak-QCD (${\cal O}(\alpha\alpha_s)$) correction has been calculated by two groups independently \cite{ewqcd1,ewqcd2}. Furthermore, the NLO and NNLO ${\cal O}(\alpha\alpha_s)$ corrections have also been computed for the final state of $\mu \Bar{\mu}H$, $i.\,e.$ including Z decays into di-muon pairs \cite{ewqcdhff}. The ${\cal O}(\alpha\alpha_s)$ correction was found to be about 1.5$\%$ of the LO result, which is significantly larger than the expected experimental accuracy of CEPC and FCC-ee.

The most important missing higher-order corrections are NNLO electroweak corrections, which are expected to contribute at the per-cent level and thus comparable or larger than the experimental precision of future Higgs factories. This letter presents the complete calculation of NNLO corrections based on two-loop electroweak diagrams with closed fermion loops. Closed fermion loops contrbutions are typically dominant because of the large top-quark Yukawa coupling and the large number of fermion flavors in the SM, which is corroborated by previous calculations \cite{Awramik:2006uz,Dubovyk:2018rlg}. To the best of our knowledge, this is the first computation of NNLO electroweak corrections to a $2\to 2$ scattering cross-section. 

The calculation is based on a semi-numerical method using a combination of dispersion relations and Feynman parametrizations, which was first introduced in Ref.~\cite{Song:2021vru} for the evaluation of two-loop double boxes. The method has been further developed to enable the treatment of UV-divergences, which occur in two-loop vertex integrals and sub-loop vertex and self-energy contributions. With our approach all relevant two-loop diagrams are reduced to at most three-dimensional numerical integrals that can be evaluated with typically 3--4 digits precision within minutes on a single CPU core [or up to a few hours when using quadruple-precision numbers for higher accuracy].

\section{Method}

\begin{figure}
\centering
\includegraphics[scale=0.5]{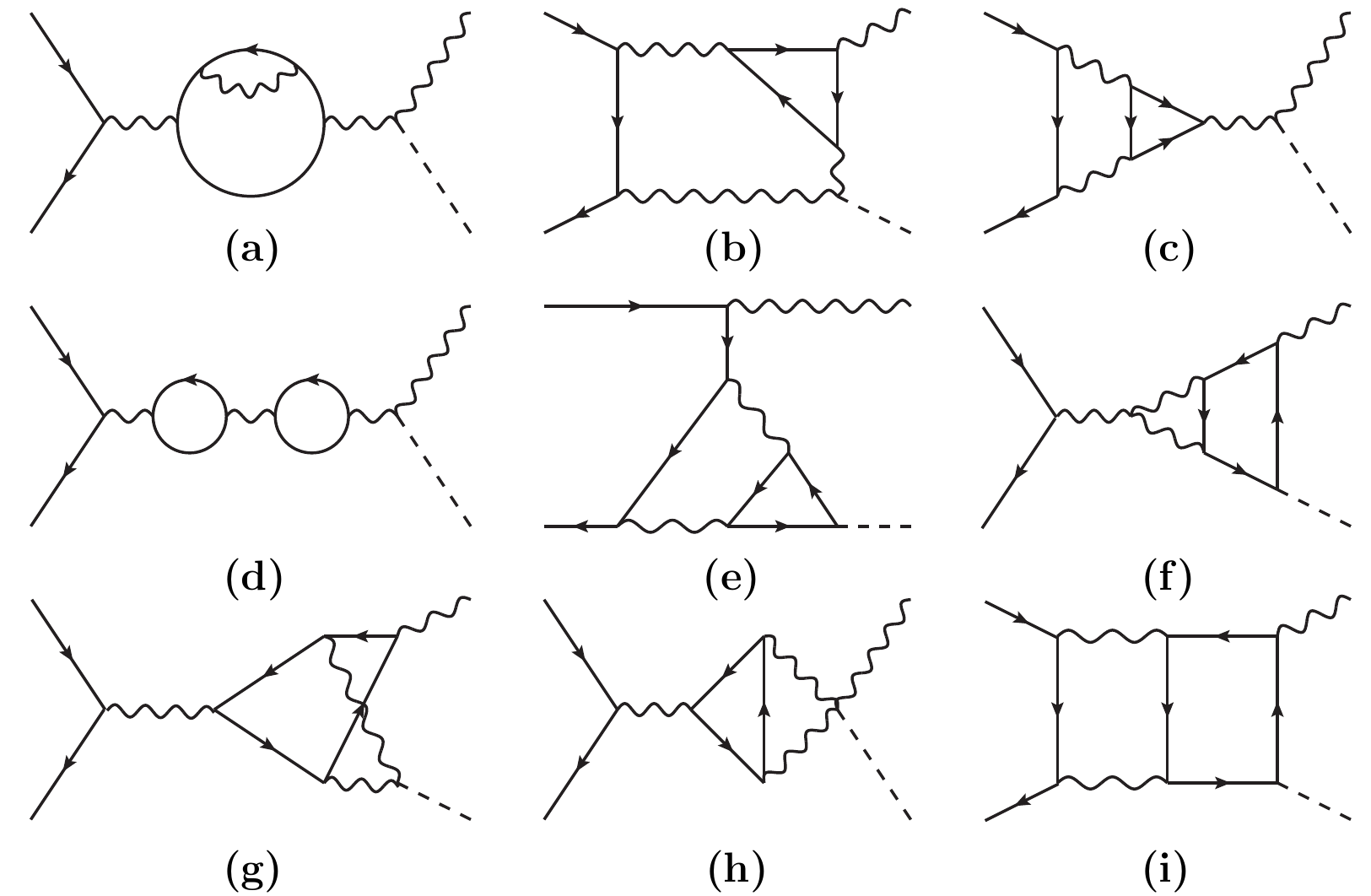}
\caption{Examples of two-loop Feynman diagrams with at least one closed fermion loop.}\label{fig:diag1}
\end{figure}

Two-loop electroweak diagrams with fermion loops can be classified into vertex, self-energy, box and reducible two-loop diagrams. Some example diagrams are shown in Fig.~\ref{fig:diag1}. Many of these diagrams are infrared (IR) and/or ultraviolet (UV) divergent. IR divergences can be spurious or physical. The former cancel in a subset of similar diagrams, but they must be regulated in individual diagrams, which we achieve by introducing a small fictitious photon mass. The physical IR divergences only emerge from initial-state QED vertex corrections, and they cancel against real photon emission contributions. However, initial-state QED corrections factorize and can be taken into account through convolution with process-independent structure functions, see $e.\,g.$ Ref.~\cite{isr2}. Therefore we omit these contribution in our calculation. Dimensional regularization is employed to regulate the UV divergence. It is worthwhile to briefly discuss the renormalization scheme and the treatment of $\gamma_5$ in $D$ dimensions.

\begin{figure}
\centering
\includegraphics[scale=0.5]{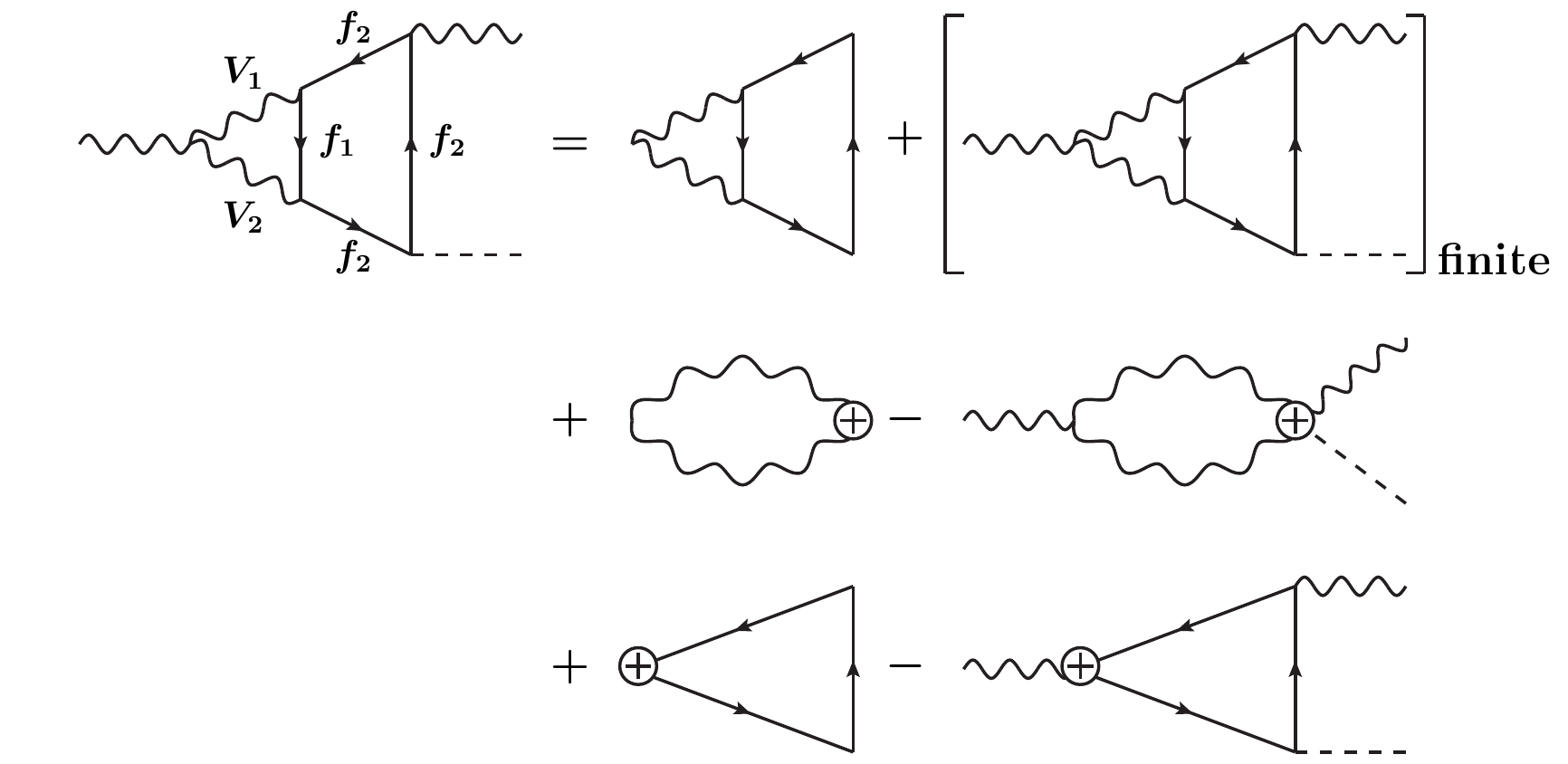}
\caption{Diagrammatical demonstration of VZH divergence separation.}\label{fig:diag2}
\end{figure}

We employed on-shell renormalization scheme for all fields, masses and electromagnetic coupling $e$. The $\alpha(0)$ scheme is used for the latter, $i.\,e.$ $e$ is normalized to its value in the Thomson limit. As a result, the final result depends on the shift $\Delta\alpha = 1-\alpha(m_Z)/\alpha(0)$, where $\alpha(\mu)$ is the running electromagnetic coupling at the sale $\mu$. More details on the renormalization parameters can be found in Ref.~\cite{Freitas:2002ja}.

The problem of $\gamma_5$ appears in the diagrams involving triangle fermion loops, which require the evaluation of $\text{tr}(\gamma^\alpha \gamma^\beta \gamma^\mu \gamma^\nu \gamma_5)$. In $D$ dimensions, the anticommutation relation $\{\gamma^\mu,\gamma_5\}$ and the trace identity $\text{tr}(\gamma^\alpha \gamma^\beta \gamma^\mu \gamma^\nu \gamma_5) = -4i\epsilon^{\alpha \beta \mu \nu}$ cannot be satisfied simultaneously. However, contributions originating from the $\epsilon$-tensor are UV finite, so they can be safely evaluated in 4 dimension. This approach has been used for example in Ref.~\cite{Awramik:2006uz,Du:2019evk}. More strategies about the treatment of $\gamma_5$ in $D$ dimensions can be found in Ref.~\cite{Jegerlehner:2000dz}. 

Now let us discuss the evaluation of the two-loop integrals in the matrix element for $e^+e^- \to ZH$. The reducible diagrams, Fig.~\ref{fig:diag1}~(d), and self-energy diagrams, Fig.~\ref{fig:diag1}~(a), can be straightforwardly computed by reducing the expressions to a set of known master integrals (MIs) \cite{Weiglein:1993hd}. The MIs have been evaluated numerically using {\sc LoopTools 2.16} \cite{looptools} for the one-loop cases and {\sc TVID 2.2} \cite{tvid} for the two-loop self-energies. The two-loop counterterms have been computed with the same approach.

For the two-loop vertex and box diagrams, we adopt the method of Ref.~\cite{Song:2021vru}, which has been extended to deal with UV-divergent diagrams. The approach uses Feynman parameters to transform one of the two sub-loops into a self-energy-type integral, which can be expressed in terms of a dispersion relation. The second sub-loop can then be solved analytically via well-known one-loop Passarino-Veltman functions. No reduction to MIs is required in this approach. The integration over the Feynman and dispersion parameters is performed numerically, resulting in at most three-dimensional integrals for two-loop vertex and box diagrams.

UV divergences need to be subtracted before carrying out the numerical integration. In general, three types of subtraction terms may be needed, two for sub-loop divergences and one more for a global (or nested) divergence. [The number of subtraction terms varies with topologies; for simpler topologies, only one term is needed.] The subtraction terms should be simple enough to be integrated analytically and then added back to the total result.

To illustrate how to subtract the UV divergences, let us take an example from the diagram shown in Fig.~\ref{fig:diag1}~(f), namely the tensor function given in Eq.~\eqref{eq:full} below. By power counting one can see that this integral has sub-loop divergences for both the $q_1$ and $q_2$ loops, as well as a global two-loop divergence. After introducing a Feynman parameter and shifting the $q_2$ momentum, one arrives at the expression in Eq.~\eqref{eq:simp1}, where $p_x = xp = x(p_z+p_h)$ and $m_x^2=(1-x)m_{V_2}^2+x m_{V_1}^2+(x^2-x)p^2$. Here $p_z$ and $p_h$ are the momenta of the final-state Z-boson and Higgs boson, respectively, whereas $p$ is the s-channel momentum. Next, the $q_2$ loop is rewritten in terms of dispersion relations. This produces a number of terms, of which only the divergent ones are explicitly shown in Eq.~\eqref{eq:simp2}. The remaining terms, denoted by $\mathcal{I}_{q_1,q_2}^\text{finite}$, are finite and do not play any role in the UV subtraction. Here $\sigma_0=(m_x+m_{f_1})^2$, and the explicit form of the dispersion kernels $\Delta B_{ij}$ can be found in Ref.~\cite{Song:2021vru}. For future reference, we introduce the symbols $\mathcal{I}_{q_1}^{ij}$ for the three $q_1$ integrals in Eq.~\eqref{eq:simp2}.
\begin{widetext}
\begin{align}
\mathcal{I} &=  \int \frac{d^D q_2}{i\pi^2} \frac{d^D q_1}{i\pi^2} 
\frac{q_2^2 q_1^2 +q_1^4}{(q_2^2-m_{V_2}^2)((q_2+p)^2-m_{V_1}^2)((q_2+q_1)^2-m_{f_1}^2)(q_1^2-m_{f_2}^2)((q_1-p_h)^2-m_{f_2}^2)((q_1-p)^2-m_{f_2}^2)} \label{eq:full} \\
&=\int_0^1 dx \int \frac{d^D q_2}{i\pi^2} \frac{d^D q_1}{i\pi^2} 
\frac{(q_2-p_x)^2 q_1^2 +q_1^4}{(q_2^2-m_x^2)^2((q_2+q_1-p_x)^2-m_{f_1}^2)(q_1^2-m_{f_2}^2)((q_1-p_h)^2-m_{f_2}^2)((q_1-p)^2-m_{f_2}^2)} \label{eq:simp1} \\
&= \int_0^1 dx \int \frac{d^D q_1}{i\pi^2} \int_{\sigma_0}^\infty d\sigma \;
\frac{\partial}{\partial m_x^2} \biggl(
\frac{q_1^4 \;\Delta B_{0}(\sigma,m_x^2,m_{f_1}^2) + q_1^2  \;\Delta B_{00}(\sigma,m_x^2,m_{f_1}^2)
+ q_1^4 \; \Delta B_{11}(\sigma,m_x^2,m_{f_1}^2)}
{(q_1^2-m_{f_2}^2)((q_1-p_h)^2-m_{f_2}^2)((q_1-p)^2-m_{f_2}^2)(\sigma-(q_1-p_x)^2)} \biggr) + \mathcal{I}_{q_1,q_2}^\text{finite}
\label{eq:simp2} 
\end{align}
\end{widetext}
The dispersion relation in~\eqref{eq:simp2} is valid only for $m_x^2 > 0$. When $m_x^2 < 0$, the dispersion relation is modified according to the following equation:
\begin{align}
    B_0(p^2,m_x^2,m_{f_1}^2) = \frac{1}{2\pi i}\int_{-\infty}^{+\infty} d\sigma \frac{B_0(\sigma,m_x^2,m_{f_1}^2)}{\sigma-p^2-i\epsilon}\,,
\end{align}
where $i\epsilon$ is a small numerical value. It must be added because it ensures all Passarino-Veltman functions are properly define for all values of $\sigma$. We have confirmed that the result is independent of $\epsilon$ as long as it is small enough. More details about dispersion relation can be found in Ref.~\cite{Song:2021vru}.

As mentioned above, $\mathcal{I}$ has a global divergence (when $q_{1,2} \to \infty$) and sub-loop divergences when either $q_1 \to \infty$ or $q_2 \to \infty$. The UV divergence from the $q_1$ loop is obtained by setting all other momenta inside $q_1$ propagators to zero. For instance, for the $q_1^4$ term in Eq.~\eqref{eq:simp1} this leads to
\begin{align}
&\mathcal{I}_{q_1}^\text{div} = \int \frac{d^D q_2}{i\pi^2} \frac{d^D q_1}{i\pi^2} \, 
\frac{1}{(q_2^2-m_{V_2}^2)((q_2-p)^2-m_{V_1}^2)} \nonumber \\
&\frac{q_1^4}{(q_1^2-m_{f_2}^2)(q_1^2-m_{f_2}^2)(q_1^2-m_{f_2}^2) (q_1^2-m_{f_1}^2)} \label{eq:divq1i} \\
&= B_0(p^2,m_{V_2}^2,m_{V_1}^2) \times [c_1 A_0(m_{f_1}^2)+c_2 A_0(m_{f_2}^2)]\,, \label{eq:divq1} 
\end{align}
where $A_0$ and $B_0$ are the usual one-loop MIs, and  the $c_i$ are functions of $m_{f_1}$,$m_{f_2}$ and the dimension $D$. The $q_2$ integral in Eq.~\eqref{eq:divq1i} can be turned into a dispersion relation and combined with Eq.~\eqref{eq:simp2} to render the $q_1$ integration finite.

The $q_2$ subloop divergence is manifested as a divergence of the $\sigma$ integral at its upper limit ($\sigma \to \infty$). In our example, the term $\partial_{m_x^2}\Delta B_{00}(\sigma,m_x^2,m_{f_1}^2)/(\sigma-(q_1-p_x)^2)$ diverges when $\sigma$ tends to infinity. However, $\partial_{m_x^2}\Delta B_{00}(\sigma,m_x^2,m_{f_1}^2) \times (\frac{1}{\sigma-(q_1-p_x)^2}-\frac{1}{\sigma-m^2})$ is UV finite, where $m^2$ can be any arbitrary value (the simplest choice is $m^2=0$). So the divergence is eliminated by subtracting the following term: 
\begin{align}
&\mathcal{I}_{q_2}^\text{div} = \int_0^1 dx \int  \frac{d^D q_1}{i\pi^2} \int_{\sigma_0}^\infty d\sigma 
\,\frac{\partial}{\partial m_x^2}
\frac{\Delta B_{00}(\sigma,m_x^2,m_{f_1}^2)}{\sigma-m^2} \nonumber \\
&\quad \times \frac{q_1^2}{(q_1^2-m_{f_2}^2)((q_1-p_h)^2-m_{f_2}^2)((q_1-p)^2-m_{f_2}^2)} \nonumber \\
&= \int_0^1 dx \; \frac{\partial}{\partial m_x^2} B_{00}(m^2,m_x^2,m_{f_1}^2) \times \mathcal{I}_{q_1}^{00}\,. \label{eq:divq2} 
\end{align}
Both factors in the last line are one-loop functions that can be computed analytically, and only the $x$ integration needs to be carried out numerically.

After subtraction of the two sub-loop divergences, one still needs to take care of the global UV divergence, which is cancelled by subtracting the same integral with all external momenta set to zero ($p_z=p_h=0$, $\therefore p_x=0$). This produces vacuum integrals, which can easily be reduced to MIs, for which analytical formulas are known \cite{2lvac}. 

As we can see in Eq~\eqref{eq:simp2}, a derivative with respect to the mass square $m_x^2$ appears in the dispersion relations for 2-loop vertex diagrams. Sometimes, for the global UV subtraction terms of diagrams with massless fermions, one can have $m_x^2=0$, which produces a singularity in the integrand. In such cases, a fictitious mass $M$ needs to be introduced for the $q_2$, $e.\,g.$
\begin{align}
\mathcal{I}^\text{div}_{q_1q_2} &= \int \frac{d^D q_2}{i\pi^2} \frac{d^D q_1}{i\pi^2}\; \frac{1}{(q_2^2-M^2)^2((q_2+q_1)^2-m_{X_1}^2)} \nonumber \\
&\qquad \frac{q_1^4}{(q_1^2-m_{X_2}^2)^3} \label{eq:vacuum} \\
& = \int \frac{d^D q_1}{i\pi^2}\, \frac{\partial B_0(q_1^2,M^2,m_{X_1}^2)}{\partial M^2}
\frac{q_1^4}{(q_1^2-m_{X_2}^2)^3}\,.
\end{align}
After adding the global subtraction term back analytically, the result does not depend on the value of $M$, which hence can be chosen arbitrarily.  
The UV subtraction process is diagrammatically demonstrated in Fig.~\ref{fig:diag2}. Diagrams without external lines correspond to vacuum diagrams. The diagrams in the second and third line are used to eliminate divergences in the subloop, which is symbolized as $\bigoplus$. They are the product of one-loop functions as shown in Eq.~\eqref{eq:divq1} and Eq.~\eqref{eq:divq2}.

The entire sequence of steps outlined above has been implemented in two independent ways to enable cross-checks. Feynman diagrams and amplitudes are generated with {\sc FeynArts}~\cite{feynarts} in both implementations. For the Lorentz and Dirac algebra, {\sc FeynCalc}~\cite{feyncalc} is employed in one implementation, and the results were cross-checked against a private code. The Feynman parametrization, construction of dispersion relations, and UV subtraction has been carried out in two independent private codes in {\sc Mathematica}. The UV-finite integrals are evaluated numerically in C++ with the help of {\sc LoopTools}~\cite{looptools} and adaptive Gauss quadrature integration, again in two separate codes, one using the integration routine from the {\sc Boost} library~\cite{boost}, and the other utilizing the {\sc Quadpack} library~\cite{quadpack}. For some cases with large numerical cancellations in the integrand, the integration was re-run using quadruple-precision numbers.
As an additional cross-check, vertex diagrams with self-energy subloops and vertex diagrams with four-point vertices (see Fig.~\ref{fig:diag1}~(h)) were also computed by reducing them to MIs and evaluating the latter with TVID~\cite{tvid}.

\section{Treatment of unstable Z boson}

Since the Z-boson width is relatively large, one needs to carefully consider the interplay of the production process $e^+e^- \to ZH$ with the Z-boson decay. For concreteness, let us focus on Z decays to the $\mu^+\mu^-$ final state. The complete all-orders matrix element for the process $e^+e^- \to \mu^+\mu^- H$ can be written as
\begin{align}
    {\cal M}_{ee\to\mu\mu H} = \Gamma_{\rm prod}\frac{1}{p_z^2-m_Z^2+\Sigma_Z(p_z^2)}\Gamma_{\rm dec} + {\cal M}_{\rm bkgd}\,, \label{matall}
\end{align}
where $\Gamma_{\rm prod}$ and $\Gamma_{\rm dec}$ are the $eeZH$ and $Z\mu\mu$ Green's function, respectively, and $\Sigma_Z$ is the Z-boson self-energy. ${\cal M}_{\rm bkgd}$ denotes any contributions to $ee\to\mu\mu H$ without a Z resonance, and it first appears at 1-loop order. For the sake of brevity, we have suppressed the Z-boson 4-vector indices in the formula above.

When expanding in perturbative orders, one encounters the problem that the individual numerator and denominator terms in Eq.~\eqref{matall} are not individually gauge invariant. A gauge-invariant prescription is given by expanding about the complex pole $s_0 = m_Z^2 - i m_Z\Gamma_Z$ of the amplitude \cite{Willenbrock:1991hu,Sirlin:1991fd,Stuart:1991xk,Veltman:1992tm},
\begin{align}
    &\begin{aligned}{\cal M}_{ee\to\mu\mu H} = 
    \,&\frac{[A]_{p_z^2=s_0}}{p_z^2-s_0} + [B]_{p_z^2=s_0} \\ &+ (p_z^2-s_0)[C]_{p_z^2=s_0} + ...\,, \end{aligned} \\ 
    &A = \frac{\Gamma_{\rm prod}\Gamma_{\rm dec}}{1+\Sigma'_Z}\,, \notag \\
    &B = \frac{\Gamma_{\rm prod}\Gamma'_{\rm dec} + \Gamma'_{\rm prod}\Gamma_{\rm dec}}{1+\Sigma'_Z} - \frac{\Gamma_{\rm prod}\Gamma_{\rm dec}\Sigma''_Z}{2(1+\Sigma'_Z)^2} + {\cal M}_{\rm bkgd}\,, \notag \\
    &C = ...\,, \notag
\end{align}
where the prime denotes derivatives $d/dp_z^2$, and for the sake of brevity we do not write the explicit form of $C$. Given that $\Gamma_Z \ll m_Z$, we can perform an additional expansion in $\Gamma_Z/m_Z$, resulting in
\begin{align}
    {\cal M}_{ee\to\mu\mu H} = \,&
    \frac{[A]_{p_z^2=m_Z^2}}{p_z^2-s_0} + \frac{p_z^2-m_Z^2}{p_z^2-s_0}[B]_{p_z^2=m_Z^2} \notag \\ &+ {\cal O}(p_z^2-m_Z^2,\Gamma_Z).
    \label{matexp}
\end{align}
Each of the terms in this equation is separately gauge invariant. Since the experimental analysis will select $\mu^+\mu^-$ pairs with invariant mass close to the Z resonance, $p_z^2 \sim m_Z^2$, the terms in the series expansion in Eq.~\eqref{matexp} decrease in numerical magnitude. Thus, if one seeks NNLO accuracy for the leading $A$ term, NLO precision is sufficient for the next-to-leading $B$ term, and LO precision for the following term (in fact, this third term is zero at tree-level and would be generated first at 1-loop order).

Focusing on the leading $A$ term, the differential cross-section, after integrating over the $\mu^+\mu^-$ angles, can be expressed as
\begin{align}
    &\frac{d^2\sigma}{d\cos\theta\,dp_z^2} = \frac{\beta}{32\pi s}\,|{\cal M}^{\rm eff}_{\rm prod}|^2 \frac{\pi^{-1}m_Z\Gamma_{Z \to \mu\mu}}{(p_z^2-m_Z^2)^2+m_Z^2\Gamma_Z^2}, \label{xsec1} \\
    &{\cal M}^{\rm eff}_{\rm prod} \equiv \frac{\Gamma_{\rm prod}}{\sqrt{1+\Sigma'_Z}}, \label{Meff} \\
    &\Gamma_{Z \to \mu\mu} \equiv \frac{|\Gamma_{\rm dec}|^2}{16\pi\, m_Z(1+\Sigma'_Z)}\,,
\end{align}
where $\beta = \sqrt{(1-m_Z^2/s-m_H^2/s)^2-4m_Z^2m_H^2/s^2}$ and $\theta$ is the scattering angle of the Higgs boson. After integration over the di-muon invariant mass, $p_z^2$, the last factor in Eq.~\eqref{xsec1} becomes approximately BR$_{Z\to\mu\mu}$, the branching fraction Z-boson into di-muon pairs. The square-root factor in \eqref{Meff} functionally takes the role of a Z-boson wavefunction renormalization. Whenever we write a squared matrix element, $|{\cal M}|^2$, it is understood to include averaging of initial-state spins and summation over final-state spins.

In this article, we present NNLO results for $\frac{d\sigma}{d\cos\theta} = \frac{\beta}{32\pi s}\,|{\cal M}^{\rm eff}_{\rm prod}|^2$. For a complete description of the process $e^+e^- \to \mu^+\mu^-H$ at this order one would also need take into account the $B$ term in \eqref{matexp} at NLO, which is a relatively straightforward one-loop calculation. We leave this combination for future work.

It is worth noting that the parametrization of the resonance according to \eqref{xsec1} leads to a definition for the Z mass and width that differs from the one that is commonly used in experimental studies. The relation between the two is given by \cite{Bardin:1988xt}
\begin{align}
m_Z = m_Z^{\rm exp}\,[1+(\Gamma_Z^{\rm exp}/m_Z^{\rm exp})^2]^{-1/2}, \\ \qquad \Gamma_Z = \Gamma_Z^{\rm exp}\,[1+(\Gamma_Z^{\rm exp}/m_Z^{\rm exp})^2]^{-1/2}.
\label{massdef}
\end{align}

\section{Results}
The following input parameters are used for the numerical evaluation:
\begin{align}
& m_W^{\rm exp} = 80.379~\text{GeV} &&\Rightarrow\quad m_W = 80.352~\text{GeV}, \nonumber \\
&m_Z^{\rm exp} = 91.1876~\text{GeV}  &&\Rightarrow\quad m_Z = 91.1535~\text{GeV}, \nonumber \\[1ex]
&m_H = 125.1~\text{GeV}, &&  m_t = 172.76~\text{GeV}, \nonumber \\ 
&\alpha^{-1} = 137.036, && \Delta \alpha = 0.059, \nonumber \\
& \sqrt{s} = 240~\text{GeV}. \end{align}
where $\sqrt{s}$ represents the center-of-mass energy, and the masses of all the other fermions are set to be 0. 

\begin{table}
\begin{ruledtabular}
\begin{tabular}{cccc}
                &  (fb)   & Contribution                     & (fb)\\
\hline
$\sigma^\text{LO}$  & 222.958 &                                  & \\
\hline
$\sigma^\text{NLO}$ & 229.893 &                              &  \\
                &         & $\mathcal{O}(\alpha_{N_f=1})$    & 21.130 \\ 
                &         & $\mathcal{O}(\alpha_{N_f=0})$    & $-$14.195 \\ 
\hline
$\sigma^\text{NNLO}$ & 231.546       &                  & \\
                &         &  $\mathcal{O}(\alpha^2_{N_f=2})$ & 1.881 \\
                &         &  $\mathcal{O}(\alpha^2_{N_f=1})$ & $-$0.226 \\ 
\end{tabular}
\end{ruledtabular}
\caption{Numerical results for the integrated cross section at LO, NLO and NNLO. Electroweak one-loop and two-loop corrections are also provided and divided according to the number of fermion loops symbolized as $N_f$.}
\label{tab:resAll}
\end{table}
\begin{figure}
\centering
\includegraphics[scale=0.6]{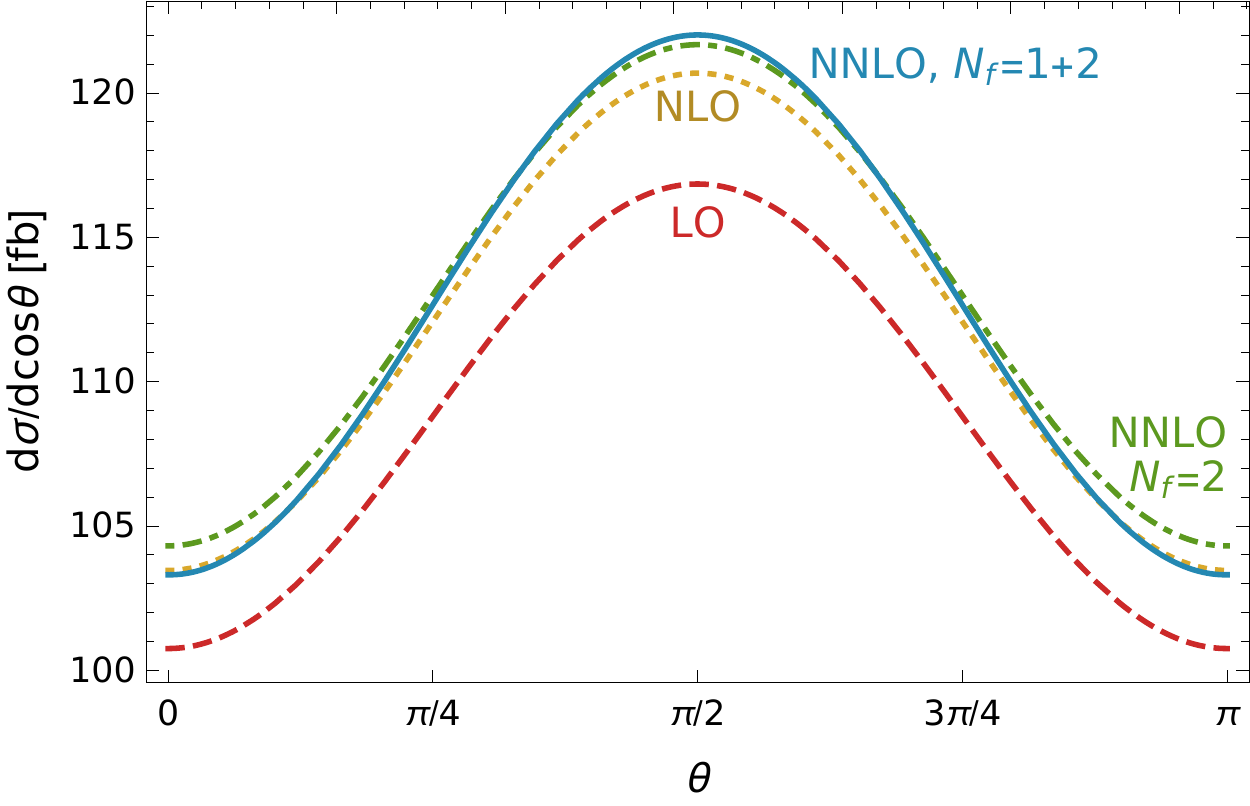}
\caption{Differential unpolarized cross section at $\sqrt{s}=240~\text{GeV}$ at LO, NLO and NNLO.}\label{fig:sigmaDiff}
\end{figure}

Table~\ref{tab:resAll} lists the results for the integrated unpolarized cross section at LO, NLO and NNLO, where corrections are further divided according to the number of closed fermion loops, denoted by $N_f$. One can see that corrections with more fermion loops dominate. At the NLO, contributions from fermionic and bosonic corrections partially cancel, resulting in an increase of $\sigma_\text{LO}$ by 3\%. The NNLO electroweak corrections turn out to be $0.7\%$ of the NLO correction, where the contribution with two fermion loops is much greater than the one with one fermion loop. This can be partially explained as a consequence of large top mass and flavor number enhancement of each fermion loop. In addition, there is an accidental numerical cancellation in the differential cross section for the $N_f=1$ contribution. This can be clearly seen in Fig.~\ref{fig:sigmaDiff}, where we plot the unpolarized differential cross section at LO, NLO and NNLO as a function of the scattering angle. The contribution due to $N_f=1$ ($i.\,e.$ the difference between the solid blue and dash-dotted green curves) is positive in the central region, $|\cos\theta| < 0.59$, and negative in the forward and backward regions, where it can reach almost $-3\%$ of the LO result.

As a consequence of this, the shape of the angular dependence in Fig.~\ref{fig:sigmaDiff} is changed slightly at NNLO in comparison to LO and NLO. This distortion mainly originates from the final-state $ZZH$/$\gamma ZH$ vertex and box diagrams.

\section{Conclusions}

Motivated by the anticipated high precision for the measurement of $\sigma(e^-e^+ \to ZH)$, in this letter we present the complete calculation of NNLO electroweak corrections with closed fermion loops. We found that they change the NLO results by 0.7\% in the $\alpha(0)$ scheme, which is comparable with the expected precision of future Higgs factories. The NNLO results can be further divided according to the number of fermion loops, and the contribution with two closed fermion loops dominates over the one with one closed fermion loop.
The calculation was made possible by a new semi-numerical technique for the evaluation of two-loop box and vertex diagrams, which could also be applied to NNLO electroweak corrections for other scattering processes.

\section*{Acknowledgments}

The authors are grateful to K.~Xie for collaboration in early stages of the project and to T.~Hahn for providing crucial improvements to {\sc LoopTools}, which are now available in version 2.16. This work has been supported in part by the National Science Foundation under grant no.~PHY-2112829.


\bibliographystyle{utcaps_mod}
\bibliography{mollerbib}

\end{document}